\documentclass[12pt,fleqn]{article}
\usepackage{times}
\usepackage[T1]{fontenc}
\normalfont
\usepackage{fourier} 
\usepackage{graphicx}% Include figure files
\usepackage{amssymb}
\usepackage{amsmath}
\usepackage{color}
\usepackage{cite}
\newcommand{\ie}{{\it i.e.}}
\newcommand{\eg}{{\it e.g.}}

\begin{document}
\baselineskip 0.6cm
\renewcommand{\thefootnote}{\#\arabic{footnote}} 
\setcounter{footnote}{0}
%\clearpage
%%%%%%%%%%%%%%%%%%%%%%%%%%%%%%%%%%%%%%%%%%%%%%%%%%%%%%%%%%%%%%%%%%%%
%%%%% ** Text ** %%%%%%%%%%%%%%%%%%%%%%%%%%%%%%%%%%%%%%%%%%%%%%%%%%%
%%%%%%%%%%%%%%%%%%%%%%%%%%%%%%%%%%%%%%%%%%%%%%%%%%%%%%%%%%%%%%%%%%%%
%
%%%%%%%%%%%%%%%%%%%%%%%%%%%%%%%%%%%%%%%%%%%%%%%%%%%%%%%%%%%%%%%%%%%%%%%
%%%%%  Title Page  %%%%%%%%%%%%%%%%%%%%%%%%%%%%%%%%%%%%%%%%%%%%%%%%%%%%
\begin{titlepage}
\begin{center}

%%%%%%%% Preprint #
\begin{flushright}
%DESY 01-142\\
\end{flushright}

\vskip 4cm

%%%%%%%% Title
{\Large \bf 
Lepton number violation by heavy Majorana neutrino in $B$ decays
}\\

\vskip 1.2cm

%%%%%%%% Authors
{\large 
Takehiko Asaka$^1$ and Hiroyuki Ishida$^2$
}

\vskip 0.4cm

%%%%%%%% Addresses
$^1${\em
  Department of Physics, Niigata University, Niigata 950-2181, Japan
}

$^2${\em
  Physics Division, National Center for Theoretical Sciences, Hsinchu 30013,
Taiwan
}

\vskip 0.2cm

%%%%%%%% Date
%(\today)
(September 20, 2016)

\vskip 2cm

%%%%%%%%%%%%%%%% Abstract %%%%%%%%%%%%%%%%%%%%%%%%%%%%%%%%%%%%%%%%%%%%%
\begin{abstract}
    Heavy Majorana neutrinos are predicted in addition to ordinary
    active neutrinos in the models with the seesaw mechanism.  We
    investigate the lepton number violation (LNV) in 
    $B$ decays induced
    by such a heavy neutrino $N$ with GeV-scale mass.  Especially, we
    consider the decay channel
    $B^+ \to \mu^+ \, N \to \mu^+ \mu^+ \pi^-$ and derive the
    sensitivity limits on the mixing angle $\Theta_\mu$ 
    by the future search experiments at Belle II 
    and in $e^+ e^-$ collisions at the Future Circular Collider
    (FCC-ee).
\end{abstract}
%%%%%%%%%%%%%%%% Abstract %%%%%%%%%%%%%%%%%%%%%%%%%%%%%%%%%%%%%%%%%%%%%
%%%%%%%%%%%%%%%%%%%%%%%%%%%%%%%%%%%%%%%%%%%%%%%%%%%%%%%%%%%%%%%%%%%%%%%
\end{center}
\end{titlepage}
%%%%%  Title Page  %%%%%%%%%%%%%%%%%%%%%%%%%%%%%%%%%%%%%%%%%%%%%%%%%%%%
%%%%%%%%%%%%%%%%%%%%%%%%%%%%%%%%%%%%%%%%%%%%%%%%%%%%%%%%%%%%%%%%%%%%%%%
%\tableofcontents

%%%%%%%%%%%%%%%%%%%%%%%%%%%%%%%%%%%%%%%%%%%%%%%%%%%%%%%%%%%%%%%%%%%%%%%
\section{Introduction}
%\noindent {\it \textbf{Introduction\,:}}~
%%%%%%%%%%%%%%%%%%%%%%%%%%%%%%%%%%%%%%%%%%%%%%%%%%%%%%%%%%%%%%%%%%%%%%%
The discovery of neutrino oscillations, showing the non-zero neutrino
masses, has opened the door to physics beyond the Standard Model~(SM).
The oscillation experiments so far have provided the rather precise
values of mass squared differences and mixing angles of active
neutrinos~\cite{Gonzalez-Garcia:2014bfa}.  There are, however, unknown
properties of active neutrinos, \ie, the ordering and the absolute values of neutrino masses, 
the violation of CP symmetry in the leptonic sector 
and the Driac or Majorana property of neutrinos.  
In addition, we do not know whether an additional particle is present
associated with the origin of neutrino masses.

Heavy neutrino is an well-motivated particle in the models of neutrino
masses.  One of the most attractive examples is the model with the
canonical seesaw
mechanism~\cite{seesaw}
where right-handed neutrinos are introduced with Majorana masses.  In
this case the mass eigenstates are three active neutrinos and heavy
neutrinos, and both neutrinos are Majorana particles.  Usually, heavy
neutrinos are considered to be much heavier than $m_W$ and even close
to the unification scale $\sim 10^{16}$~GeV.  Such heavy particles are
attractive since they can also account for the baryon asymmetry of the
universe (BAU) via leptogenesis~\cite{Fukugita:1986hr}.  

On the other hand, heavy neutrinos with masses below $m_W$ are also
attractive.  Even in this case the seesaw mechanism is still effective
by requiring the suppressed Yukawa coupling constants of neutrinos.
Furthermore, the BAU can be explained by using the different
mechanism~\cite{Akhmedov:1998qx,Asaka:2005pn}.  Heavy neutrinos with
$\sim 100$ MeV are interesting for the supernova
explosion~\cite{Fuller:2009zz}.  If its mass is around keV scale, it
can be a candidate for the dark matter~\cite{Dodelson:1993je}.  Futher
it may explain the origin of pulsar
velocities~\cite{pulsar}.
(See, for example, Ref.~\cite{Kusenko:2009up} for astrophysics of
heavy neutrinos.)  Therefore, heavy neutrinos which are lighter than
the electroweak scale are also well-motivated particles beyond the SM.
Interestingly, such particles can be tested in terrestrial
experiments~\cite{terrestrial}.

If neutrinos are Majorana particles, the lepton number of the SM
Lagrangian is broken.  In this case there appear various phenomena
which are absent in the SM.  The contribution from heavy Majorana
neutrino can be significant depending on its mass and mixing.  The
well-known example is the neutrinoless double beta decay
$(Z,A) \to (Z+2,A) + 2 e^-$. See, for example, a recent
review~\cite{Pas:2015eia} and references therein.  When the mass is of
the order of 0.1--1~GeV, the contribution from heavy Majorana neutrino
can be significant to alter the prediction of the rate solely from
active neutrinos.

The LNV process $e^- e^- \to W^- W^-$ (called as the inverse
neutrinoless double beta decay~\cite{Rizzo:1982kn}) is another
interesting possibility to test the Majorana property of heavy neutrino.  
Various aspects of this process have been investigated so far~\cite{i0nbb}. 
%See, for example, the recent analyses in Refs.~\cite{i0nbb} and the references therein.
It is a good target of the future lepton colliders such as the
International Linear Collider (ILC)~\cite{Baer:2013cma} and the
Compact Linear Collider (CLIC)~\cite{Accomando:2004sz}.

Another example is the rare decay of meson like
$M^+ \to \ell^+ \ell'{}^+ M'{}^-$ where $M$ and $M'$ are mesons and
$\ell$ and $\ell'$ are charged leptons with the same
charge~\cite{Ng:1978ij,Abad:1984gh,Littenberg:1991ek,Dib:2000wm,Ali:2001gsa,Atre:2005eb,Cvetic:2010rw,Canetti:2014dka,Milanes:2016rzr,Cvetic:2016fbv}.  See the current experimental limits on these processes in
Refs.~\cite{terrestrial,Agashe:2014kda}.  Heavy Majorana neutrino with 
an appropriate mass gives a sizable contribution to these processes, and
its mixing receives the upper bounds from the experimental data.

In this paper we discuss the LNV decay of $B$ mesons 
induced by heavy Majorana neutrino with GeV-scale mass.
In particular, we study the testability of the mode $B^+ \to \mu^+ \mu^+ \pi^-$
by the future experiments.  The expected limits 
on the mixing of heavy neutrino by 
Belle II~\cite{SuperKEKB} and the $e^+ e^-$ collisions 
on $Z$-pole at the future circular collider (FCC-ee)~\cite{FCCee}
will be presented.

%%%%%%%%%%%%%%%%%%%%%%%%%%%%%%%%%%%%%%%%%%%%%%%%%%%%%%%%%%%%%%%%%%%%%%%
\section{Heavy Majorana neutrino}
%\noindent {\it \textbf{Heavy Majorana neutrino\,:}}~
%%%%%%%%%%%%%%%%%%%%%%%%%%%%%%%%%%%%%%%%%%%%%%%%%%%%%%%%%%%%%%%%%%%%%%%
We consider a heavy Majorana neutrino $N$ with mass $M_N \sim $ GeV
which mixes with ordinary left-handed neutrinos $\nu_{L \alpha}$
($\alpha = e, \mu, \tau$) as
\begin{align}
  \nu_{L \alpha} = U_{\alpha i} \, \nu_i + \Theta_{\alpha } \, N \,,
\end{align}
where $U_{\alpha i}$ is the PMNS mixing matrix of active neutrinos
$\nu_i$ ($i=1,2,3$).  In this case $N$ has the weak gauge interactions which are suppressed by the mixing $\Theta_\alpha$.
Here we discuss only one heavy neutrino for simplicity, but the extension
to the case with more heavy neutrinos is straightforward by replacing
$\Theta_\alpha N$ with $\sum_I \Theta_{\alpha I} N_I$.

If heavy neutrinos provide the tiny neutrino masses through the seesaw mechanism,
the masses and mixings of heavy neutrinos
must satisfy a certain relation to explain the experimental results
of the neutrino oscillations.  However, we do not specify 
the origin of $N$ to make a general argument and consider
$M_N$ and $\Theta_\alpha$ as free parameters in this analysis.

It is possible to test directly heavy neutrino $N$ by various
experiments because of the smallness of its mass.  Since there is no
signal of this particle, the upper bounds on the mixing
$|\Theta_\alpha|$ are imposed from various experiments depending on
its mass~\cite{terrestrial}.  
It is then important to search it by future experiments at the first step.
Furthermore, not only the discovery but also 
the detail study is crucial to reveal the properties of $N$.

\begin{figure}[t]
  \centerline{
  \includegraphics[width=10cm]{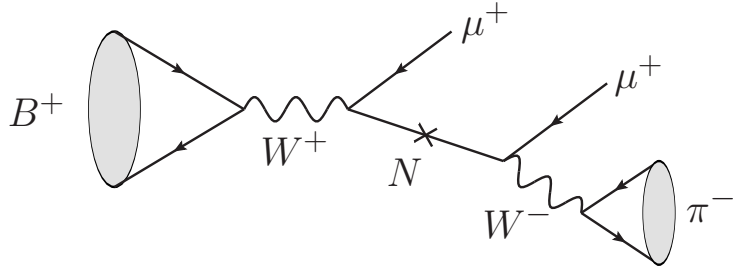}
  }%
  \caption{
    LNV decay process of charged $B$ meson.
  }
  \label{Fig:Bdec}
\end{figure}

In the present analysis we consider the experimental test 
for the LNV to show the Majorana property of $N$. 
Especially, we focus on the LNV decay of $B$ meson 
as a concrete example % 
\footnote{In this analysis we discuss only the decay into two muons,
  but the extension to the decays into the like sign leptons with
  other flavors is straightforward.}
\begin{align}
  \label{eq:LNVB}
  B^+ \to \mu^+ \, N \to \mu^+ \,  \mu^+ \, \pi^- \,,
\end{align}
which is mediated by the on-shell $N$ as shown in Fig.~\ref{Fig:Bdec}.
%%%%%%%%%%%%%%%%%%%%%%%%%%%%%%%%%%%%%%%%%%%%%%%%%%%%%%%%%%%%%%%%%%%%%%
%%%%% ** Figure ** %%%%%%%%%%%%%%%%%%%%%%%%%%%%%%%%%%%%%%%%%%%%%%%%%%%
%%%%%%%%%%%%%%%%%%%%%%%%%%%%%%%%%%%%%%%%%%%%%%%%%%%%%%%%%%%%%%%%%%%%%%
Notice that there is also the charge conjugated process which is implicit from now on.
From the kinematical reason we restrict ourselves to the mass region
\begin{align}
  m_B - m_\mu > M_N > m_{\pi} + m_\mu \,.
\end{align}
In the process (\ref{eq:LNVB}) 
the production rate of $N$ is proportional to $|\Theta_\mu|^2$
and the decay rate is also proportional to $|\Theta_\mu|^2$,
and then the LNV signal is induced as the $|\Theta_\mu|^4$ effect.
This process has been discussed as an interesting target 
for Belle and LHCb experiments~\cite{terrestrial,Cvetic:2010rw,Canetti:2014dka,%
Milanes:2016rzr,Cvetic:2016fbv}.

\begin{figure}[ht]
  \centerline{
  \includegraphics[width=17cm]{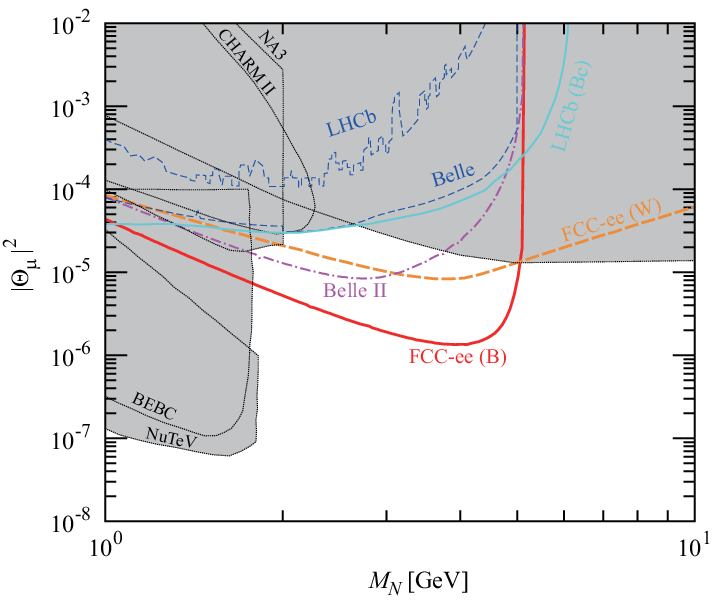}%
  }%
  \caption{ 
    The sensitivity limits on $|\Theta_\mu|^2$ from the LNV decay
    $B^+ \to \mu^+ \mu^+ \pi^-$ due to heavy neutrino at Belle II
    with $N_B=5 \times 10^{10}$ (magenta dot-dashed line) and at
    FCC-ee with $N_Z = 10^{13}$ (red solid line).  
    The orange long-dashed line is the limit 
    from $W^+ \to \mu^+ \mu^+ \pi^-$ at FCC-ee with 
    $N_W = 2 \times 10^{8}$.
    For comparision we also show the limit
    from the LNV decays $B^+_c \to \mu^+ \mu^+ \pi^+$ at LHCb for LHC run
    3~\cite{Milanes:2016rzr}  (cyan solid line). 
    The blue dashed lines are the upper
    bounds from the LNV $B$ decays by LHCb~\cite{Aaij:2014aba} and
    Belle~\cite{Liventsev:2013zz}.  The gray region is excluded by
    search experiments: DELPHI~\cite{Abreu:1996pa},
    NA3~\cite{Badier:1986xz}, CHARM II~\cite{Vilain:1994vg},
    BEBC~\cite{CooperSarkar:1985nh}, and NuTeV~\cite{Vaitaitis:1999wq}.
  }
  \label{Fig:UB}
\end{figure}

The recent results of the search for $B^+ \to \mu^+ \mu^+ \pi^-$
are obtained 
by Belle~\cite{Liventsev:2013zz} and LHCb~\cite{Aaij:2014aba}.
(See also Ref.~\cite{Shuve:2016muy} for the revision of the LHCb limit.)
They presented the upper bounds on the mixing
$|\Theta_\mu|^2$ as shown in Fig.~\ref{Fig:UB}.
In the same figure we also present
various constraints on heavy neutrino 
which are from Ref.~\cite{terrestrial}.
It is found that these bounds on $|\Theta_\mu|^2$ 
are weaker than other constraints on heavy neutrino which are 
applicable to both Dirac and Majorana cases.

The future prospect of the LHCb search for the LNV decays of $B$ and
$B_c$ mesons including (\ref{eq:LNVB}) has been discussed in
Ref.~\cite{Milanes:2016rzr}.  The sensitivity on the mixing 
by using the mode $B_c^+ \to \mu^+ \mu^+ \pi^-$ at LHC run 3, 
which is better than that of (\ref{eq:LNVB}), 
is also shown in Fig.~\ref{Fig:UB}.
In the present analysis, we then investigate the 
search for the process (\ref{eq:LNVB}) 
at Belle II and FCC-ee.

%%%%%%%%%%%%%%%%%%%%%%%%%%%%%%%%%%%%%%%%%%%%%%%%%%%%%%%%%%%%%%%%%%%%%%
%%%%% ** Figure ** %%%%%%%%%%%%%%%%%%%%%%%%%%%%%%%%%%%%%%%%%%%%%%%%%%%
%%%%%%%%%%%%%%%%%%%%%%%%%%%%%%%%%%%%%%%%%%%%%%%%%%%%%%%%%%%%%%%%%%%%%%

%%%%%%%%%%%%%%%%%%%%%%%%%%%%%%%%%%%%%%%%%%%%%%%%%%%%%%%%%%%%%%%%%%%%%%%
\section{Search at Belle II}
%\noindent {\it \textbf{Search at Belle II\,:}}~
%%%%%%%%%%%%%%%%%%%%%%%%%%%%%%%%%%%%%%%%%%%%%%%%%%%%%%%%%%%%%%%%%%%%%%%

Let us first consider the search for the LNV decay 
of $B^+$ shown in Eq.~(\ref{eq:LNVB}) at Belle~II~\cite{SuperKEKB},
where $5 \times 10^{10}$ pairs of $B$ mesons (at 50~ab$^{-1}$)
are planned to be produced.
In this analysis we take 
the number of $B^+$ as $N_{B} = 5 \times 10^{10}$
and the energy as $E_{B} = m_{B^\pm}$ since the velocity of produced $B^\pm$'s is low enough.  Let us then estimate the expected number
of the signal events below.

First, the partial decay rate of $B^+ \to \mu^+ N$ is given by
\begin{align}
  \Gamma (B^+ \to \mu^+ \, N)
  &=
    \frac{G_F^2 \, f_{B^\pm}^2 \, m_{B^\pm}^3}{8 \pi}  \,
    |V_{ub}|^2 \, |\Theta_{\mu}|^2 
    \left[ r_\mu^2 + r_N^2 - (r_\mu^2 - r_N^2)^2 \right]
    \sqrt{  1 - 2( r_\mu^2 + r_N^2 ) + (r_\mu^2 - r_N^2)^2 } \,,
\end{align}
where $f_{B^\pm}$ is the decay constant, $V_{ub}$ is the CKM element, and 
\begin{align}
  r_\mu = \frac{m_\mu}{m_{B^\pm}} \,,~~~~~
  r_N = \frac{M_N}{m_{B^\pm}} \,.
\end{align}
Notice that the rate is enhanced by $M_N^2/m_\mu^2$ 
for $M_N \gg m_\mu$ because of the helicity suppression effect 
of this process.  
%The branching ratio of $B^+ \to \mu^+ N$ is estimated as
In oder to avoid the uncertainty in $f_B$ and $V_{ub}$ the branching ratio of $B^+ \to \mu^+ N$  is estimated as
\begin{align}
  Br (B^+ \to \mu^+ N) = 
  \frac{\Gamma (B^+ \to \mu^+ N)}{\Gamma ( B^+ \to \tau^+ \nu_\tau)}
  \times Br( B^+ \to \tau^+ \nu_\tau) \,,
\end{align}
where the branching ratio of $B^+ \to \tau^+ \nu_\tau$
is $Br( B^+ \to \tau^+ \nu_\tau) = (1.14 \pm 0.27) \times 10^{-4}$~\cite{Agashe:2014kda}.  In order to estimate the number of the signal events
the energy distribution of $N$ in $B^+ \to \mu^+ N$ is important since it determines the decay length of $N \to \mu^+ \pi^-$.
In the present case due to the two-body decay at rest 
it is simply given by
\begin{eqnarray}
  E_N = \frac{m_{B^\pm}^2 + M_N^2 - m_\mu^2}{2 m_{B^\pm}} \,.
\end{eqnarray}
The number of the signal events is then 
\begin{eqnarray}
  \label{eq:Nevent1}
  N_{\rm event} 
  &=
    2 \, N_{B^+} \, Br (B^+ \to \mu^+ N) \,
    P( N \to \mu^+ \pi^- ; E_N, L_{\rm det} ) \,,
\end{eqnarray}
where $P( N \to \mu^+ \pi^- ; E_N, L_{\rm det} )$ 
is the probability that the signal decay $N \to \mu^+ \pi^-$
occurs inside the detector, which is given by
\begin{align}
  \label{eq:P}
  P( N \to \mu^+ \pi^- ; E_N, L_{\rm det} )
  = \frac{\Gamma (N \to \mu^+ \pi^-)}{\Gamma_N}
  \left[ 1 - \exp \left( - \frac{M_N \Gamma_N L_{\rm det}}{E_N} \right) \right] \,,
\end{align}
where $\Gamma_N$ is the total decay rate of $N$.
We calculate $\Gamma_N$
for the case when $\Theta_\mu \neq 0$ and
$\Theta_e = \Theta_\tau =0$
taking into account the possible 
decay channels by using the expressions for the partial rates
in Ref.~\cite{Gorbunov:2007ak}.
On the other hand, the partial rate of $N \to \mu^+ \pi^-$ 
is given by
\begin{align}
  \Gamma (N\to \mu^+ \pi^-) 
  &=
    \frac{1}{16 \pi}
    |\Theta_\mu|^2 
    |V_{ud}|^2 G_F^2 f_{\pi^\pm}^2 M_N^3
    \left[
    \left( 1 - \frac{m_\mu^2}{M_N^2} \right)^2
    - 
    \frac{m_{\pi^\pm}^2}{M_N^2}
    \left( 1 + \frac{m_\mu^2}{M_N^2} \right)
    \right]
    \nonumber \\
  &\times
    \left[
    1 - 2 \frac{m_{\pi^\pm}^2 + m_\mu^2}{M_N^2}
    + \frac{(m_{\pi^\pm}^2 - m_\mu^2)^2}{M_N^4}
    \right]^{1/2} \,.
\end{align}
Here we take $m_{\pi^\pm}=139.6$~MeV, 
$f_{\pi^\pm}=130.4$~MeV and $|V_{ud}|=0.9743$~\cite{Agashe:2014kda}.
The typical size of the detector is denoted by
$L_{\rm det}$ and we take it as $L_{\rm det}=1.5$~m 
for Belle II detector for simplicity.
Note that the factor 2 in Eq.~(\ref{eq:Nevent1})
represents the contribution from the charge conjugate 
process of (\ref{eq:LNVB}).

We assume that there is no background event and 
the sensitivity limit on $|\Theta_\mu|^2$ at 95~\% C.L.
is obtained from $N_{\rm event}=3.09$~\cite{Feldman:1997qc}.
The result is shown in Fig.~\ref{Fig:UB}.
It is seen that Belle II can probe the LNV effect 
by heavy neutrino with $M_N \simeq 2$--3~GeV 
and $|\Theta_\mu|^2 = {\cal O}(10^{-5})$
which is consistent with various experimental constraints.%
\footnote{This issue has also been discussed in Ref.~\cite{Cvetic:2016fbv}.
Although they have not presented the quantitative estimate
of the limit, their qualitative result is consistent with ours.}
Interestingly, the sensitivity is better than the test
of $B_c^+ \to \mu^+ \mu^+ \pi^-$ at LHCb for LHC run 3~\cite{Milanes:2016rzr}.

%%%%%%%%%%%%%%%%%%%%%%%%%%%%%%%%%%%%%%%%%%%%%%%%%%%%%%%%%%%%%%%%%%%%%%%
\section{Search at FCC-ee}
%\noindent {\it \textbf{Search at FCC-ee\,:}}~
%%%%%%%%%%%%%%%%%%%%%%%%%%%%%%%%%%%%%%%%%%%%%%%%%%%%%%%%%%%%%%%%%%%%%%%
Next, we turn to consider the search at the future plan,
the $e^+ \, e^-$ collisions at the Future Circular Collider (FCC-ee).
It is planned to produce $10^{12}$-$10^{13}$ $Z$ bosons at
the $Z$-pole $\sqrt{s}=m_Z$.  The direct search for heavy neutrino
at FCC-ee has been discussed in Ref.~\cite{Blondel:2014bra}.
The method there cannot clarify whether heavy neutrino is a
Dirac or Majorana particle.  Here we shall discuss the sensitivity
of the LNV process~(\ref{eq:LNVB}) aiming to test the Majorana 
property of heavy neutrino.

The number of $B^+$ in $Z$ decays is estimated as
\begin{align}
  N_{B^+} = N_Z \times Br ( Z \to b \bar b) \times f_u \,,
\end{align}
where $N_Z$ is the number of $Z$ produced at FCC-ee,
and $N_Z = 10^{13}$ is assumed in the present analysis.
$Br (Z \to b \bar{b})=0.1512$~\cite{Agashe:2014kda} is the branching ratio of
$Z \to b \bar{b}$ and $f_u = 0.410$~\cite{Amhis:2014hma} is the
fraction of $B^+$ from $\bar b$ quark in $Z$ decay.
It is then found that $N_{B^+} = 6.20 \times 10^{-2} \, N_Z$ 
is much larger than that in the case of Belle II, 
from which we can expect the much better sensitive at FCC-ee.
Although the produced $B^+$'s have the energy distribution peaked at $E_{B^+} \sim 40$~GeV (see, \eg, Ref.~\cite{Ackerstaff:1998zf}), 
we shall set
\begin{align}
  E_{B^+} = \frac{m_Z}{2} \,,
\end{align}
for simplicity.  In this case the distribution of the energy of $N$ in $B^+ \to \mu^+ N$ is flat as
\begin{align}
  \frac{1}{\Gamma_{B^+ \to \mu^+ N}} \frac{d \Gamma_{B^+ \to \mu^+ N}}{d E_N}
  = \frac{1}{p_{B^+} \beta_f} \,,
\end{align}
for the energy range $E_N^+ \ge E_N \ge E_N^-$.
Here $p_{B^+}= \sqrt{E_{B^+}^2 - m_{B^\pm}^2}$ and 
\begin{align}
  \beta_f 
  &= \sqrt{ 1 - \frac{2 (M_N^2+m_\mu^2)}{m_{B^\pm}^2}
  + \frac{(M_N^2 - m_\mu^2)^2}{m_{B^\pm}^4}} \, ,
  \\
  E_N^\pm
  &=
    \frac{4 (m_{B^\pm}^2 + M_N^2 - m_\mu^2) E_{B^+} 
    \pm 4 p_{B^+} m_{B^\pm}^2 \beta_f}{8 m_{B^\pm}^2} \,.
\end{align}
The number of the signal events~(\ref{eq:LNVB}) is then estimated as 
\begin{align}
  N_{\rm event} 
  &=
    2 \, \int_{E_N^-}^{E_N^+} dE_N
    N_{B^+} \, Br (B^+ \to \mu^+ N) \,
    \frac{1}{p_{B^+} \beta_1} \,
    P( N \to \mu^+ \pi^- ; E_N, L_{\rm det} ) \,.
\end{align}
Now we take $L_{\rm det}=2$~m
for the probability $P( N \to \mu^+ \pi^- ; E_N, L_{\rm det} )$ 
in Eq.~(\ref{eq:P}).

In Fig.~\ref{Fig:UB} we also show the sensitivity limit on the mixing
$|\Theta_\mu|^2$ from the LNV decay $B^+ \to \mu^+ \mu^+ \pi^-$ at
FCC-ee with $N_Z =10^{13}$.  As in the previous case
we assumed no background event and estimate the limit
from $N_{\rm event}=3.09$. 
We can see that FCC-ee improves greatly the sensitivity
compared with those of Belle II and LHCb for LHC run 3.
For heavy Majorana neutrino with $M_N \simeq 4$~ GeV
the mixing $|\Theta_\mu|^2 \gtrsim 10^{-6}$ 
can be probed.  Thus, FCC-ee can offer the significant test
of the LNV by heavy Majorana neutrino.

One might think that the LNV signal might be boosted for $N$ produced
in $B_c$ mesons, since the partial rate of $B_c^+ \to N + \mu$ receives
a milder suppression factor $|V_{cb}|^2 = 1.69 \times 10^{-3}$ rather
than $|V_{ub}|^2=1.71 \times 10^{-5}$~\cite{Agashe:2014kda}.  The
production of $B_c$ in $Z$ decays, however, is hard and the branching
ratio is
$Br (Z \to B_c^+ + b + \bar c) = (2.04-3.33) \times
10^{-5}$~\cite{Bcprod}.
Thus, the LNV events through $B_c$ meson is smaller than those through $B$
and then we shall neglect it in the present analysis. 
It is, however, an interesting target for LHCb experiment as discussed 
in Ref.~\cite{Milanes:2016rzr}.  See also Fig.~\ref{Fig:UB}.

We should mention that FCC-ee offers another promising test of the LNV
induced by heavy Majorana neutrino.%
\footnote{ The Majorana property of heavy neutrino may also be probed
  from $e^+ e^- \to N \nu \to \ell q \bar q' \nu$ by using the angular
  distribution between $N$ and the incoming
  $e^-$~\cite{delAguila:2005pin}.  In addition, the LNV process like
  $e^+ e^- \to N e^\pm W^\mp \to \ell^\pm W^\pm e^\pm W^\mp$ leading
  to the same-sign dilepton with four hadronic jets is also an
  interesting target~\cite{Banerjee:2015gca}.  } 
It is planned to produce more than $2 \times 10^8$ $W$ pairs at the
center-of-mass energy at the $WW$ threshold and
above~\cite{Gomez-Ceballos:2013zzn}.  In this case the LNV decay
$W^+ \to \ell^+ N \to \ell^+ \ell'{}^+ \pi^-$ can be tested.\footnote{
The LNV decay of $W$ at LHC has been discussed in Refs.~\cite{LNVW}.}  
The sensitivity limit on $|\Theta_\mu|^2$ by using this mode is also shown
in Fig.~\ref{Fig:UB}.  It is found that the sensitivity by using
$B^+ \to \mu^+ \mu^+ \pi^-$ is better than this for the parameter
range in which constraints are avoided.

%%%%%%%%%%%%%%%%%%%%%%%%%%%%%%%%%%%%%%%%%%%%%%%%%%%%%%%%%%%%%%%%%%%%%%%
\section{Summary}
%\noindent {\it \textbf{Summary\,:}}~
%%%%%%%%%%%%%%%%%%%%%%%%%%%%%%%%%%%%%%%%%%%%%%%%%%%%%%%%%%%%%%%%%%%%%%%
We have discussed the LNV decay of $B$ meson,
$B^+ \to \mu^+ \mu^+ \pi^-$, induced by heavy Majorana neutrino.
In particular we have estimated the sensitivity limits
on the mixing $|\Theta_\mu|^2$ by the experimental searches
at Belle II and at FCC-ee (at $Z$-pole).
These facilities can probe the parameter region 
in which the various experimental constraints on heavy neutrino
are avoided.  Thus, the LNV $B$ decay 
is a significant and promising target for the LNV,
which is complementary to the neutrinoless double beta decay.

%%%%%%%%%%%%%%%%%%%%%%%%%%%%%%%%%%%%%%%%%%%%%%%%%%%%%%%%%%%%%%%%%%%%
%%%%%%%%%%%%%%%%%%%%%%%%%%%%%%%%%%%%%%%%%%%%%%%%%%%%%%%%%%%%%%%%%%%%
\section*{Acknowledgments}
%%%%%%%%%%%%%%%%%%%%%%%%%%%%%%%%%%%%%%%%%%%%%%%%%%%%%%%%%%%%%%%%%%%%
%%%%%%%%%%%%%%%%%%%%%%%%%%%%%%%%%%%%%%%%%%%%%%%%%%%%%%%%%%%%%%%%%%%%
The work of T.A. was partially supported by JSPS KAKENHI
Grant Numbers 15H01031and 25400249.
T.A. thanks the Yukawa Institute for Theoretical Physics at Kyoto
University, where this work was initiated during the YITP-S-16-01 on
"The 44th Hokuriku Spring School".

%%%%%%%%%%%%%%%%%%%%%%%%%%%%%%%%%%%%%%%%%%%%%%%%%%%%%%%%%%%%%%%%%%%%
%%%%%%%%%%%%%%%%%%%%%%%%%%%%%%%%%%%%%%%%%%%%%%%%%%%%%%%%%%%%%%%%%%%%
%%%%% ** Reference ** %%%%%%%%%%%%%%%%%%%%%%%%%%%%%%%%%%%%%%%%%%%%%%
%%%%%%%%%%%%%%%%%%%%%%%%%%%%%%%%%%%%%%%%%%%%%%%%%%%%%%%%%%%%%%%%%%%%
%%%%%%%%%%%%%%%%%%%%%%%%%%%%%%%%%%%%%%%%%%%%%%%%%%%%%%%%%%%%%%%%%%%%

%%%%%%%%%%%%%%%%%%%%%%%%%%%%%%%%%%%%%%%%%%%%%%%%%%%%%%%%%%%%%%%%%%%%
%%%%%%%%%%%%%%%%%%%%%%%%%%%%%%%%%%%%%%%%%%%%%%%%%%%%%%%%%%%%%%%%%%%%
%%%%%%%%%%%%%%%%%%%%%%%%%%%%%%%%%%%%%%%%%%%%%%%%%%%%%%%%%%%%%%%%%%%%
%%%%%%%%%%%%%%%%%%%%%%%%%%%%%%%%%%%%%%%%%%%%%%%%%%%%%%%%%%%%%%%%%%%%
\end{document}